\documentclass[english,prd,showpacs,aps,superscriptaddress]{revtex4-1}
\usepackage[latin9]{inputenc}
\usepackage{amsmath}
\usepackage{amssymb}
\usepackage{graphicx}
\usepackage{esint}

\makeatletter
 
 \@ifundefined{textcolor}{}
 {%
   \definecolor{BLACK}{gray}{0}
   \definecolor{WHITE}{gray}{1}
   \definecolor{RED}{rgb}{1,0,0}
   \definecolor{GREEN}{rgb}{0,1,0}
   \definecolor{BLUE}{rgb}{0,0,1}
   \definecolor{CYAN}{cmyk}{1,0,0,0}
   \definecolor{MAGENTA}{cmyk}{0,1,0,0}
   \definecolor{YELLOW}{cmyk}{0,0,1,0}
 }

\usepackage{mathrsfs}\usepackage{latexsym}

\makeatother

\usepackage{babel}
\begin{document}

\title{Phase structures of holographic screen themodynamics}

\author{Wei-Jian Jiang}

\email{jwj@nit.zju.edu.cn}

\affiliation{Zhejiang Institute of Modern Physics, Zhejiang University, Hangzhou
310027, China}

\affiliation{Ningbo Institute of Technology, Zhejiang University, Ningbo 315100,
China}

\author{Yi-Xin Chen}

\email{yixinchenzimp@zju.edu.cn}

\thanks{Corresponding author}

\affiliation{Zhejiang Institute of Modern Physics, Zhejiang University, Hangzhou
310027, China}

\author{Jian-Long Li}

\email{jlli@zimp.zju.edu.cn}

\affiliation{Zhejiang Institute of Modern Physics, Zhejiang University, Hangzhou
310027, China}
\begin{abstract}
Holographic screens are the generalization of the event horizon of
a black hole in entropic force scheme, which are defined by setting
Newton potential $\phi$ constant, \textit{i. e.} $e^{2\phi}=c=$const.
By demonstrating that the integrated first law of thermodynamics is
equivalent to the ($r-r$) component of Einstein equations, We strengthen
the correspondence between thermodynamics and gravity. We show that
there are not only the first law of thermodynamics, but also kinds
of phase transitions of holographic screens. These phase transitions
are characterized by the discontinuity of their heat capacities. In
(n+1) dimensional Reissner-Nordstr\"{o}m-anti-de Sitter (RN-AdS)
spacetime, we analyze three kinds of phase transitions, which are
of the holographic screens with $Q=0$ (charge), constant $\Phi$
(electrostatic potential) and non-zero constant $Q$. In the $Q=0$
case, only the holographic screens with $0\le c<1$ can undergo phase
transition. In the constant $\Phi$ case, the constraints become as
$0\le c+16\tilde{\Gamma}^{2}\Phi^{2}<1$, where $\tilde{\Gamma}$
is a dimensional dependent parameter. By verifying the Ehrenfest equations,
we show that the phase transitions in this case are all second order
phase transitions. In the constant $Q$ case, there might be two,
or one, or no phase transitions of holographic screens, depending
on the values of $Q$ and $c$.
\end{abstract}

\pacs{04.70.Dy, 04.20.Cv, 04.40.Nr, 04.50.-h}

\maketitle

\section{Introduction}

Since 1970s, black hole becomes a new research window for quantum
theories of gravity. Although it is a characteristic solution of general
relativity, surprisingly, black hole behaves like a thermodynamical
system in the semi-classical methods. Namely, when considering the
quantum feature, a black hole is found to have the Hawking temperature
\cite{Hawking1975}, Bekenstein-Hawking entropy \cite{Bekenstein1973},
and four laws of black hole thermodynamics \cite{bardeen1973four}.
Furthermore, the black holes defined in Anti-de Sitter (AdS) space
are found to have thermodynamical phase transitions \cite{hawking1983thermodynamics}.
These quantum features are verified and reproduced by various discussions
in various manners. The lesson we learn from this window is summarized
as that, the degrees of freedom of a quantum gravity system can be
described by the degrees of freedom on its boundary \cite{tHooft:1993gx,Susskind:1994vu}.
As a result, the holographic principle is proposed.

Both gravity and thermodynamics are universal laws in nature \cite{Zhao:2010qw}.
Since gravity is so intensely correlated with black hole thermodynamics,
people started to envision that there might be only one universal
law, and gravity is somehow originated in thermodynamics. In order
to test this idea, Jacobson \cite{PhysRevLett.75.1260} showed that
the Einstein field equations of gravity can be deduced from the first
law of thermodynamics on local Rindler horizon. By introducing a maximization
principle of Wald entropy, Padmanabhan \cite{Padmanabhan:2007en,Padmanabhan:2009vy}
also obtained Einstein's equations on null-like hypersurface. Verlinde
\cite{Verlinde:2010hp} proposed that gravity might not be a fundamental
interaction but an entropic force. He obtained Newton's law of gravity
from his assumptions, and left Einstein's equations as an unresolved
problem. The idea of entropic force also has some applications\cite{Cai:2010kp,Cai:2010zw}.
 Besides, there are also other propositions pursuing the idea that
the nature of gravity is thermodynamics of unknown microstructure
of spacetime. This situation reminds people with Boltzmann's speculation
on classical thermodynamics before the quantum theory of atoms was
discovered \cite{Padmanabhan:2010xh}: \textit{if you can heat it,
it has microstructure.}

In this paper, we follow the entropic force scenario to investigate
the phase transitions of holographic screens. Holographic screen is
proposed by Verlinde, and is mathematically defined by Chen \textit{et
al. }in \cite{Chen:2010ay}. It is assumed to be the elemental unit
of space, which preserves Newton potential $\phi$ as constant. For
example, the event horizon of a black hole is a special holographic
screen. Thus, when generalizing the black hole temperature, energy,
holographic degrees of freedom to the holographic screens, Verlinde
argued that there might be a thermodynamical law of holographic screens,
in which gravity is emergent. In \cite{Chen:2010ay}, Chen \textit{et
al. }studied the thermodynamics of the holographic screen, and illustrated
the idea about how gravity and space are emergent from lower-dimensional
holographic screen. In \cite{Tian:2010gn}, Tian \textit{et al. }studied
it in higher-deimensional spacetime and in Gauss-Bonnet gravity. These
previous works are inspiring, but preliminary. There are still some
gaps between black hole thermodynamics and the holographic screen
thermodynamics. For instance, how do we reproduce the Einstein equations,
or a part of them in holographic screen thermodynamics? Is there any
phase transitions of holographic screens? These are what we concern
in this paper.

We briefly review the holographic screen thermodynamics in an (n+1)
dimensional RN-AdS spacetime. If we start from the integrated form
of the first law of the holographic screen thermodynamics, we can
obtain the ($r-r$) component of the Einstein equations. It extends
the correspondence of thermodynamics and gravity from null-like surfaces
to time-like ones.

When we set $Q=0$, the RN-AdS holographic screens become the AdS
ones. In these cases, we show that there are phase transitions similar
to Hawking-Page phase transition of Schwarzschild AdS black hole.
The existence of phase transition picks out a special group of holographic
screens, which are with $0\le c<1$. There are unstable phases for
holographic screens with smaller mass, smaller entropy and negative
heat capacity, while stable phases for those with larger mass, larger
entropy and positive heat capacity. The temperature and mass of the
phase transition point are determined by $c$.

Moreover, we consider the phase transition of holographic screen with
constant electrostatic potential $\Phi$ in RN-AdS metric. The holographic
screens must satisfy $0\le(c+16\tilde{\Gamma}^{2}\Phi^{2})<1$ to
have phase transition. There are unstable phases with smaller mass,
larger charge and negative heat capacity, while stable phases with
larger mass, smaller charge and positive heat capacity. Using Ehrenfest
scheme, we find that these phase transitions of holographic screens
are all second order phase transitions.

The phase transitions of holographic screens with constant $Q$ are
more intricate. A mutual limit of the holographic screen minimal mass
arises because of the non-extremal condition for RN-AdS black hole.
As a result, different holographic screens will have different numbers
of phases, and the features of the phase structure are determined
by the values of $Q$ and$c$. For $n=3$ and $l=1$, we show clearly
the regeons of $(Q,c)$ for each specific phase structure. And there
is also a critical point for the holographic screen with $c<0.623$.

There is also a remarkable feature we should mention here. Since there
are several definitions of quasi-local energy in general relativity,
generally speaking, the idea about the thermodynamical formalism of
gravity has two main variations. The first variation is based on the
ADM mass energy and Komar energy \cite{Padmanabhan:2010xh,Verlinde:2010hp,Pan:2010eu,Tian:2010uy,Chen:2010ay}.
They were widely used in black hole thermodynamics since the discovery
of black hole thermodynamics. Although the ADM mass energy is a global
definition of the energy stored on the whole space, it can be proved
as an effective quasi-local energy transformed from Komar energy stored
on part of the space under a Legendre transformation \cite{Chen:2010ay}.
In this case, the first law of the black hole thermodynamics obeys
the corresponding Maxwell relations, and every thermodynamical function
can be expressed as a state function of the thermodynamical parameters,
such as $M$, $J$ and $Q$ \cite{Banerjee2009}. Hence, it is a fine
analogy of classical thermodynamics. The second variation is based
on other energy definitions, such as Misner-Sharp energy \cite{PhysRevD.81.061501,PhysRevD.81.084012,Tian:2010gn},
Brown-York energy \cite{Brown,Gu:2010wv} and etc. Although they can
not be related one another via Legendre transformations, they are
well-defined in time dependent cases. As a result, the second variation
tried to extend the black hole thermodynamics from static background
to general cases, \textit{e.g.} the thermodynamics of FLRW universes
\cite{PhysRevD.81.061501}. In summary, these two variations both
have their own advantages. We will focus our discussion on the first
variation.

This paper is organized as follows. In Sec. 2, we give a brief review
on holographic screen and the correspondence of the first law of thermodynamics
and the Einstein equations. In Sec. 3, we analyze the holographic
screen phase transitions with $Q=0$, constant $\Phi$ and constant
$Q$ respectively, and we introduce the Ehrenfest scheme to analyze
the phase transition of holographic screen with constant $\Phi$.
Sec. 4 presents the conclusion and discussion.

\section{Thermodynamics of holographic screen}

Previously, the formulae of holographic screen thermodynamics is investigated
in 4-dimensional spherical static spacetime in \cite{Chen:2010ay}.
Since we are interested in phase transitions of holographic screen
in (n+1)-dimensional Reissner-Nordstr\"{o}m-anti-de Sitter metric,
we first generalize the formulae in \cite{Chen:2010ay} to the higher-dimensional
RN-AdS case. In the context, we use the units $G=\hbar=c=k_{B}=1$.

The (n+1)-dimensional RN-AdS metric ($n\ge3$) is \cite{Niu:2011tb}
\begin{equation}
ds^{2}=-f(r)dt^{2}+\frac{dr^{2}}{f(r)}+r^{2}d\Omega_{n-1}^{2},\label{RNAdS}
\end{equation}
 where 
\begin{equation}
f(r)=1-\frac{8\tilde{\Gamma}M}{r^{n-2}}+\frac{Q^{2}}{r^{2n-4}}-\frac{2\Lambda r^{2}}{n(n-1)}.\label{f}
\end{equation}
 Here, $d\Omega_{n-1}$ is the line element of a unit (n-1)-sphere.
$M$ is the ADM mass energy. $Q$ is the charge parameter which is
different from the electrostatic charge by a dimension-dependent factor
\cite{Banerjee:2011au}. $\Lambda=-n(n-1)/(2l^{2})$ is the negative
cosmological constant, and $l$ is the AdS radius. $\tilde{\Gamma}$
is a short notation for $\Gamma(\frac{n}{2})/[(n-1)\pi^{(n/2)-1}]$.

The Newton potential is defined by the global time-like Killing vector
$\xi^{\mu}=(1,0,...,0)$,

\begin{equation}
\phi=\frac{1}{2}\ln{(-\xi^{2})}=\frac{1}{2}\ln{f(r)}.\label{Newton}
\end{equation}
We can define the holographic screen as an equipotential hypersurface
with constant Newton potential,

\begin{equation}
f(r)=e^{2\phi}=c.\label{screen}
\end{equation}
The other parameters vary according to different thermal configurations.
Here $c$ is a constant, satisfying $c\ge0$ for RN-AdS metric. When
$c=0$, the holographic screen is identical to the event horizon of
RN-AdS black hole. Thus the holographic screen thermodynamics can
be viewed as a generalization of the black hole thermodynamics on
null-like hypersurface to time-like hypersurface.

For some specific values of $M$, $Q$, $\Lambda$ and $c$, we can
solve $r_{s}$ from Eq. (\ref{screen}) which is the radial parameter
of the holographic screen. As a result, the mass of a holographic
screen can be written as

\begin{equation}
M=\frac{1}{8\tilde{\Gamma}}\bigg\{(1-c)r_{s}^{n-2}+\frac{Q^{2}}{r_{s}^{n-2}}-\frac{2\Lambda r_{s}^{n}}{n(n-1)}\bigg\}.\label{M}
\end{equation}

The temperature of the holographic screen, which is proposed by Verlinde
\cite{Verlinde:2010hp} as the redshifted Unruh temperature detected
by an accelerating observer, is determined by the Newton potential,
\textit{i. e.}

\begin{eqnarray}
T & = & \frac{1}{2\pi}e^{\phi}\sqrt{\nabla^{\mu}\phi\nabla_{\mu}\phi}=\frac{1}{4\pi}\bigg|\frac{\partial f(r_{s})}{\partial r_{s}}\bigg|\nonumber \\
 & = & \frac{1}{4\pi r_{s}}\bigg\{(n-2)\Big[(1-c)-\frac{Q^{2}}{r_{s}^{2(n-2)}}\Big]-\frac{2\Lambda r_{s}^{2}}{n-1}\bigg\}.\label{T}
\end{eqnarray}

If the energy and temperature of holographic screen are determined,
the entropy is uniquely defined via \cite{Chen:2010ay}

\begin{equation}
S=-4\pi\int\big(\frac{\partial f(r_{s})}{\partial M}\big)^{-1}dr_{s}=\frac{\pi r_{s}^{n-1}}{2(n-1)\tilde{\Gamma}}.\label{S}
\end{equation}
In \cite{Chen:2010ay}, the authors argued that the entropy is only
defined on the (n-1) dimensional holographic screen. It is also interpreted
as that the area of holographic screen is proportional to and determined
by the holographic screen entropy, i.e. $A_{n-1}=4S$. We take this
definition as an assumption when exploring the phase transition of
holographic screen.

The corresponding electrostatic potential $\Phi$ to the charge parameter
$Q$ is 
\begin{equation}
\Phi=\frac{\partial M}{\partial Q}=\frac{1}{4\tilde{\Gamma}}\frac{Q}{r_{s}^{n-2}}.\label{Phi}
\end{equation}

The conjugate potential of $\Lambda$ is

\begin{equation}
\Theta=\frac{\partial M}{\partial\Lambda}=-\frac{r_{s}^{n}}{4\tilde{\Gamma}n(n-1)}.\label{Theta}
\end{equation}

As a result, it is possible to address the first law of holographic
screen thermodynamics as

\begin{equation}
dM=TdS+\Phi dQ+\Theta d\Lambda.\label{firstlaw}
\end{equation}
By integrating Eq. (\ref{firstlaw}), we get the integrated form of
the first law, which is in other words the generalized Smarr formula,

\begin{equation}
(n-2)M=(n-1)TS+(n-2)\Phi Q-2\Theta\Lambda.\label{Smarr}
\end{equation}

In \cite{Chen:2010ay}, the authors conjectured that we can also take
thermodynamics as our starting point, and define gravity from Eqs.
(\ref{firstlaw}) and (\ref{Smarr}). We illustrate this idea as follows.
Specifically, if we set

\begin{equation}
r_{s}=\bigg(\frac{2(n-1)\tilde{\Gamma}S}{\pi}\bigg)^{1/(n-1)},
\end{equation}
 
\begin{equation}
f(r_{s})=1-4\tilde{\Gamma}(2M-\Phi Q-2\Theta\Lambda)r_{s}^{2-n},
\end{equation}
and $\partial f(r_{s})/\partial r_{s}=4\pi T$, Eq. (\ref{Smarr})
is equivalent to

\begin{equation}
r_{s}\frac{\partial f}{\partial r_{s}}-(n-2)(1-f)=\frac{16\pi P}{n-1}r_{s}^{2}.\label{Grr}
\end{equation}
We recognize Eq. (\ref{Grr}) as the $(r-r)$ component of Einstein
equations in the neighborhood of holographic screen, where $P$ is
the effective radial pressure contributed by the negative cosmological
constant and the Maxwell field, which is

\begin{equation}
P=-\frac{\Lambda}{8\pi}-\frac{(n-1)(n-2)\tilde{\Gamma}}{4\pi}\frac{\Phi Q}{r_{s}^{n}}.\label{P}
\end{equation}

Thus we have got an illustration for the emergence of gravity from
(n-1) dimensional holographic screen thermodynamics. This is the first
illustration about obtaining the Einstein equations from thermodynamics
on time-like surfaces. Although the observation above sounds plausible,
and can be regarded as a generalization of black hole thermodynamics,
the other aspects and characteristics of holographic screen thermodynamics
are still unknown. In next section, we will study the phase structure
of holographic screens in (n+1)-dimensional RN-AdS spacetime.

\section{Phase structure of holographic screen}

The phase structure of a black hole is always analyzed where the heat
capacity $C$ is divergent. Take a Schwarzschild AdS black hole as
an example, it is found that given the AdS radius $l$, black hole
has a minimal temperature $T_{min}(l)$. The heat capacity of the
black hole at this temperature is divergent. Increasing the mass of
the black hole, its heat capacity will turn to a finite positive value,
and its temperature will rise. Decreasing the mass of the black hole
, its heat capacity will turn to a finite negative value, and its
temperature will also rise. So the black hole will become unstable.
This is the typical Hawking-Page phase transition\cite{hawking1983thermodynamics,Nicolini:2011dp}.
For RN-AdS black hole, there are also phase transitions with constant
$Q$ and constant $\Phi$ \cite{Banerjee:2011raa,Banerjee:2011au,Niu:2011tb,Majhi:2012fz}.

For holographic screen in RN-AdS metric, things get a little more
complicated, because there are a congruence holographic screens with
different Newton potential $c$ and thermodynamical entities in the
same spacetime. We should investigate the phase structure's dependence
on $c$ as well as the holographic screen phase transition. In order
to get a quick view of holographic screen phase transition, first
we study the $Q=0$ case. And then, we analyze the constant $\Phi$
and $Q$ cases respectively.

\subsection{The $Q$=0 case}

When $Q=0$, the metric becomes the same as the Schwarzschild AdS
case. The mass, temperature and heat capacity of a holographic screen
become,

\begin{equation}
M=\frac{1}{8\tilde{\Gamma}}\bigg\{(1-c)r_{s}^{n-2}-\frac{2\Lambda r_{s}^{n}}{n(n-1)}\bigg\}.\label{M_AdS}
\end{equation}

\begin{equation}
T=\frac{1}{4\pi r_{s}}\bigg\{(n-2)(1-c)-\frac{2\Lambda r_{s}^{2}}{n-1}\bigg\}.\label{T_AdS}
\end{equation}

\begin{equation}
C=\frac{\partial M}{\partial T}=\frac{\pi r_{s}^{n-1}}{2\tilde{\Gamma}}\frac{(n-2)(1-c)-\frac{2\Lambda}{n-1}r_{s}^{2}}{-(n-2)(1-c)-\frac{2\Lambda}{n-1}r_{s}^{2}}.\label{C_AdS}
\end{equation}

\begin{figure}[h]
\centering \includegraphics[width=0.4\textwidth]{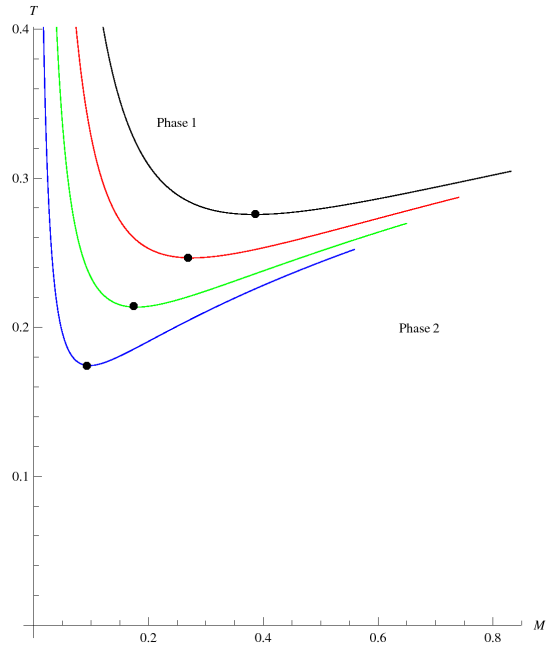} \caption{\label{f1}Temperature ($T$) vs. mass ($M$) for holographic screens,
with $n=3$ and $l=1$. The values of $c$ for black, red, green,
blue lines are 0, 0.2, 0.4, 0.6, respectively.}
\end{figure}

\begin{figure}[h]
\centering \includegraphics[width=0.4\textwidth]{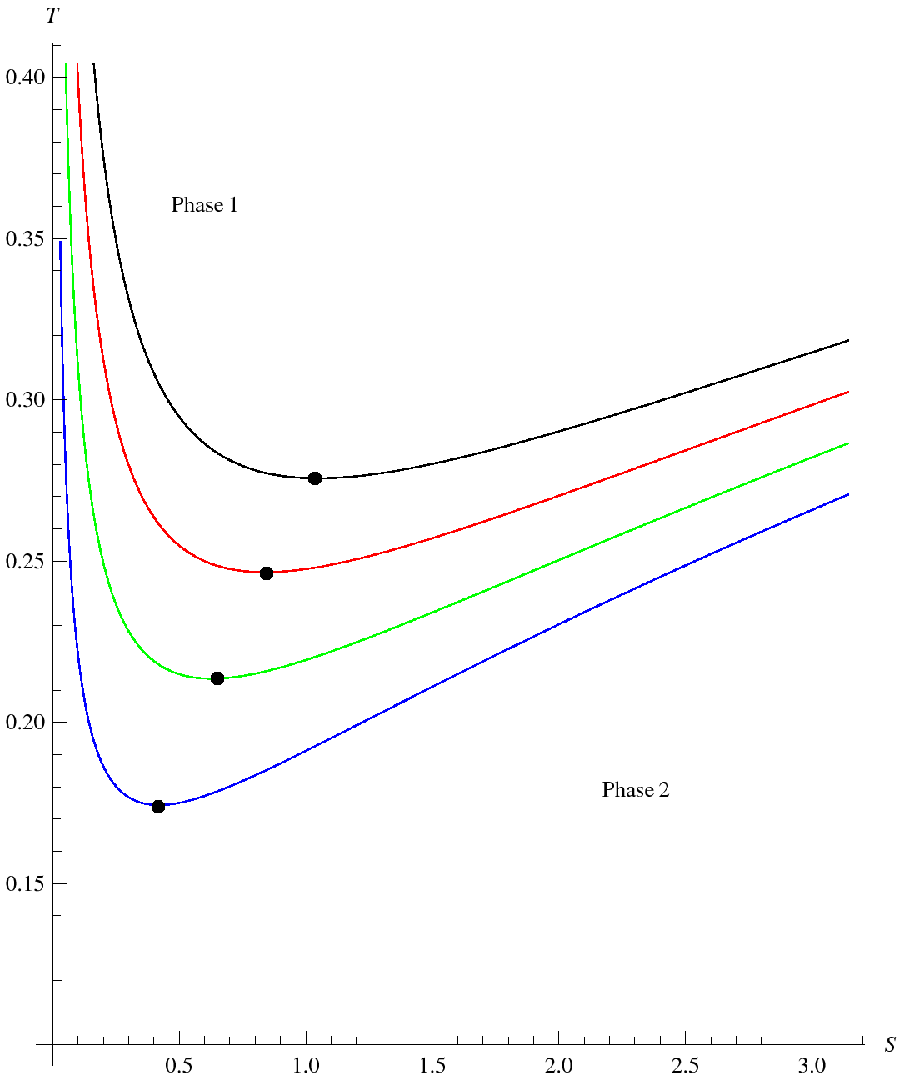} \caption{\label{f2}Temperature ($T$) vs. entropy ($S$) for holographic screens,
with $n=3$ and $l=1$. The values of $c$ for black, red, green,
blue lines are 0, 0.2, 0.4, 0.6, respectively.}
\end{figure}

\begin{figure}[h]
\centering \includegraphics[width=0.4\textwidth]{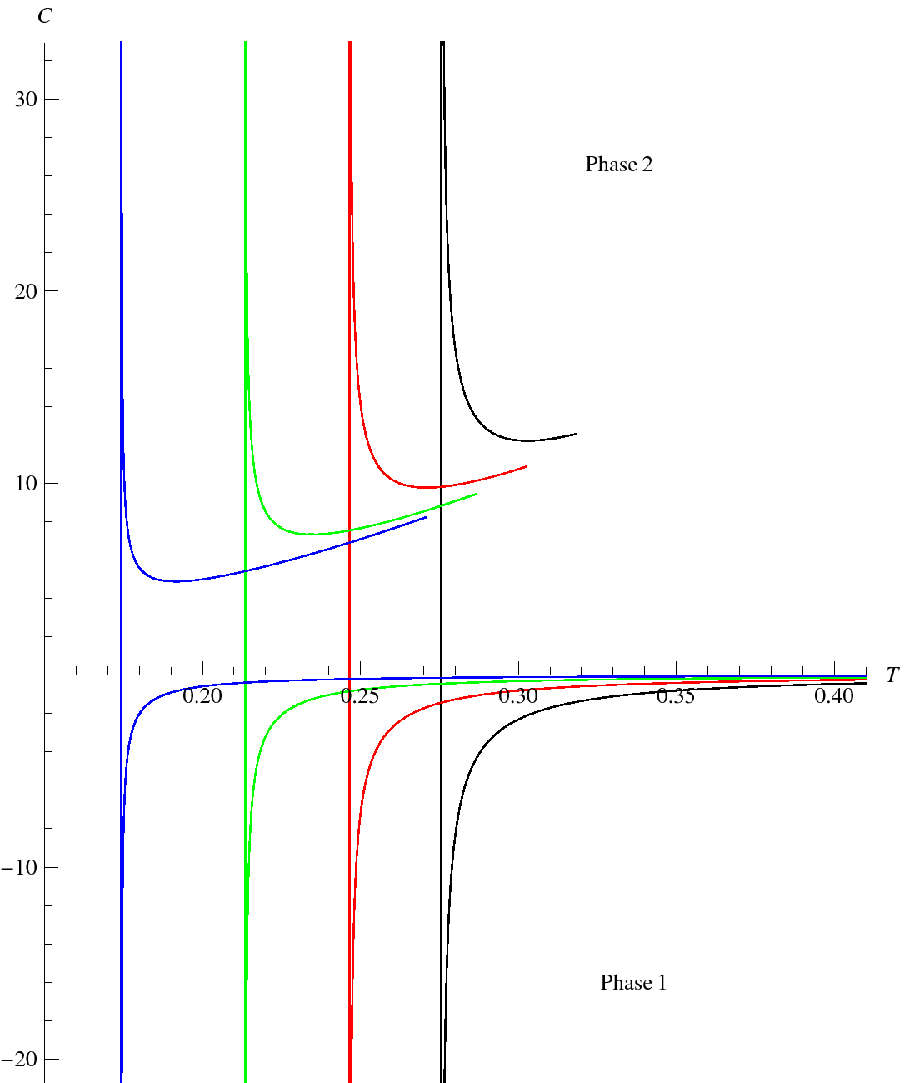} \caption{\label{f3}Heat capacity ($C$) vs. temperature ($T$) for holographic
screens, with $n=3$ and $l=1$. The values of $c$ for black, red,
green, blue lines are 0, 0.2, 0.4, 0.6, respectively.}
\end{figure}

\begin{figure}[h]
\centering \includegraphics[width=0.4\textwidth]{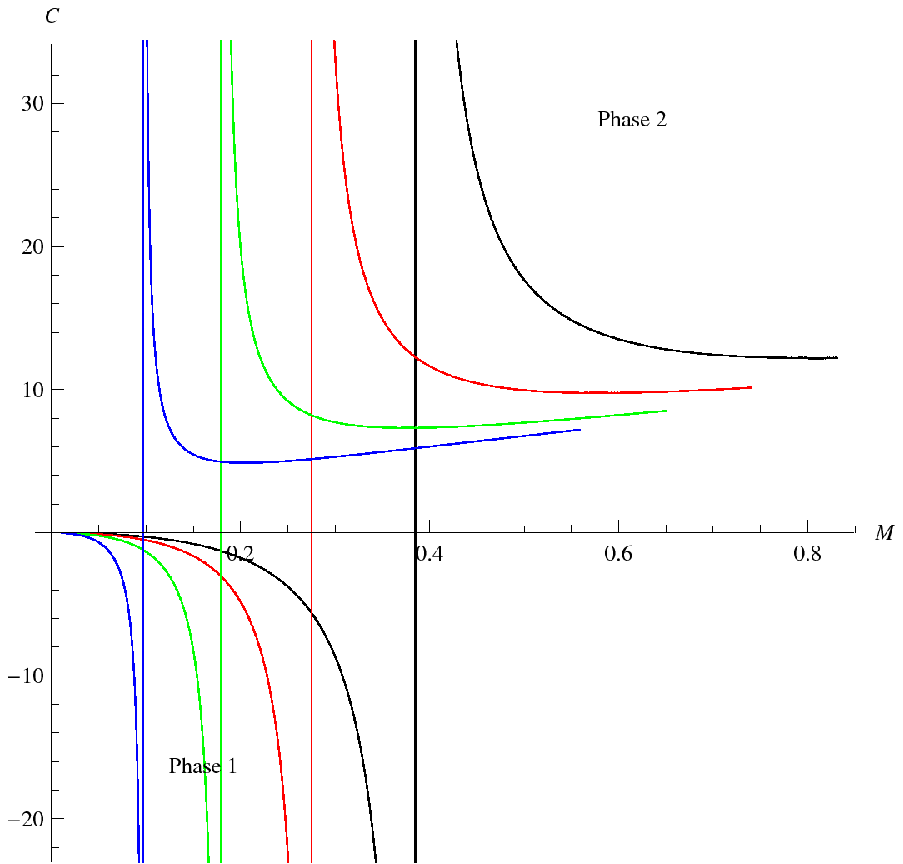} \caption{\label{f4}Heat capacity ($C$) vs. mass ($M$) for holographic screens,
with $n=3$ and $l=1$. The values of $c$ for black, red, green,
blue lines are 0, 0.2, 0.4, 0.6, respectively.}
\end{figure}

Since $\Lambda$ is a constant and $c$ is a parameter which specifies
the holographic screen, the only independent variable in the phase
structure is the mass $M$. In Figs. \ref{f1}, \ref{f2}, \ref{f3},
\ref{f4}, we present the $T$-$M$, $T$-$S$, $C$-$T$, $C$-$M$
diagrams of holographic screens with different $c$. (Without loss
of generality, we set $n=3$ and $l=1$ in all the figures. The values
of $c$ for black, red, green, blue lines are 0, 0.2, 0.4, 0.6, respectively.)
Each holographic screen with specific $c$ in Figs. \ref{f1}-\ref{f4}
undergoes a phase transition at the point $C\rightarrow\infty$. (We
marked these turning points in Figs.\ref{f1}, \ref{f2}.) By setting
the denominator in Eq. (\ref{C_AdS}) as 0, we get $r_{s}$ at the
phase tansition point,

\begin{equation}
r_{pt}^{2}=\frac{(n-2)(n-1)(1-c)}{-2\Lambda}.\label{eq:rpt_AdS}
\end{equation}

From Figs. \ref{f1}, \ref{f2}, \ref{f3}, \ref{f4}, we notice the
following features. There are two phases of every holographic screen,
which are separated by the turning point. When the holographic screen
is in phase 1, it has smaller mass, smaller entropy and negative heat
capacity. In contrary, in phase 2 it has larger mass, larger entropy
and positive heat capacity. The phase transition occurs between phase
1 and phase 2, and at the phase transition point the heat capacity
divergent.

According to Eq. (\ref{C_AdS}), the phase transition exists when
the Newton potential $c$ satisfies $0\le c<1$. There is no phase
transition for holographic screens with $c\ge1$. It indicates that
the holographic screens with $0\le c<1$ are a special group of the
whole congruence.

The temperature of the holographic screen at the transition point,
$T_{pt}$, can be deduced from Eq. (\ref{eq:rpt_AdS}) and Eq. (\ref{T_AdS}),
\begin{equation}
T_{pt}=\frac{1}{2\pi}\sqrt{\frac{-2\Lambda(n-2)(1-c)}{n-1}}.\label{eq:Tpt}
\end{equation}
When $0\le c<1$, the larger $c$ is, the lower $T_{pt}$ becomes,
which also can be seen in Fig. \ref{f1}, \ref{f2}. $T_{pt}$ is
also the minimum temperature of the corresponding holographic screen.

The mass of the holographic screen at the transition point, $M_{pt}$,
can be deduced from Eq. (\ref{eq:rpt_AdS}) and Eq. (\ref{M_AdS}),
\begin{equation}
M_{pt}=\frac{n-1}{4\tilde{\Gamma}n}(\frac{(n-1)(n-2)}{-2\Lambda})^{(n-2)/2}(1-c)^{n/2}.\label{eq:Mpt}
\end{equation}
The holographic screen with mass smaller than $M_{pt}$ is in phase
1, and with mass larger than $M_{pt}$ is in phase 2. When $0\le c<1$,
the larger $c$ is, the smaller $M_{pt}$ becomes, as shown in Fig.
\ref{f1}, \ref{f2}. It also indicates the holographic screens will
undergo phase transitions in some order. 

We can demonstrate the above idea by a gedanken experiment as in Figs.
\ref{f1} and \ref{f4}. At first, there is nothing but an (n+1) dimensional
AdS spacetime with $n=3$ and $l=1$. Then we add a mass shell with
a little mass $\Delta m$ in the spacetime. As a result, the holographic
screen with $c_{0}=1-3(\frac{\Delta m}{2})^{2/3}$ is undergoing its
phase transition from phase 1 to phase 2. At the moment, the holographic
screens with $c_{0}<c<1$ are already in phase 2, while those with
$0\le c<c_{0}$ are still in phase 1. As we add more and more mass
into the spacetime, $c_{0}$ becomes smaller and smaller, which means
more holographic screens have transited from phase 1 to phase 2. When
the mass reaches $M=\frac{2}{3}\sqrt{\frac{1}{3}}$, the black hole
event horizon as a special holographic screen with $c=0$ is undergoing
its phase transition, and the other screens are all in phase 2. In
summary, when we gradually increase the mass of the spacetime, the
screen with larger $c$ will undergo its phase transition earlier.

From Fig. \ref{f4} we see, when $c$ gets larger, the curve becomes
more cliffy, which means the heat capacity decreases (in phase 1)
or grows (in phase 2) more rapidly as $M$ comes closer to $M_{pt}$.
It indicates the holographic screen phase transition with larger $c$
is more acute about the change of $M$.

\subsection{Phase transition with constant $\Phi$}

We consider the phase transition of charged holographic screen with
constant $\Phi$. The mass, temperature and heat capacity of holographic
screen become,

\begin{equation}
M=\frac{1}{8\tilde{\Gamma}}\bigg\{(1-c+16\tilde{\Gamma}^{2}\Phi^{2})r_{s}^{n-2}-\frac{2\Lambda r_{s}^{n}}{n(n-1)}\bigg\}.\label{M_RNAdS}
\end{equation}

\begin{equation}
T=\frac{1}{4\pi r_{s}}\bigg\{(n-2)(1-c-16\tilde{\Gamma}^{2}\Phi^{2})-\frac{2\Lambda r_{s}^{2}}{n-1}\bigg\}.\label{T_RNAdS}
\end{equation}

\begin{align}
C_{\Phi} & =(\frac{\partial M}{\partial T})_{\Phi}=T(\frac{\partial S}{\partial T})_{\Phi}\label{C_RNAdS1}\\
 & =\frac{\pi r_{s}^{n-1}}{2\tilde{\Gamma}}\frac{(n-2)(1-c-16\tilde{\Gamma}^{2}\Phi^{2})-\frac{2\Lambda}{n-1}r_{s}^{2}}{-(n-2)(1-c-16\tilde{\Gamma}^{2}\Phi^{2})-\frac{2\Lambda}{n-1}r_{s}^{2}}.\label{eq:C_RNAdS2}
\end{align}

In this case, there are two independent parameters, the Newton potential
$c$ and the electrostatic potential $\Phi$. When we increase the
mass of a holographic screen with constant $c$, from Eq. (\ref{M_RNAdS})
and Eq. (\ref{Phi}), we can see that, in order to preserve $\Phi$
constant, we must also decrease its charge parameter $Q$. As a result,
there is only one independent variable $M$ in analysing the phase
structure.

\begin{figure}[h]
\centering \includegraphics[width=0.4\textwidth]{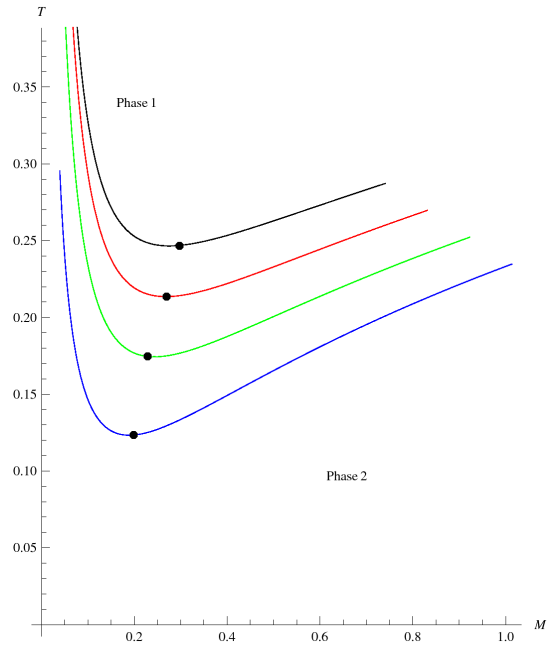} \caption{\label{f5}Temperature ($T$) vs. mass ($M$) for holographic screens
with constant $\Phi$, with $n=3$, $l=1$ and $c=0.2$. The values
of $\Phi^{2}$ for black, red, green, blue lines are 0, 0.2, 0.4,
0.6, respectively.}
\end{figure}

\begin{figure}[h]
\centering \includegraphics[width=0.4\textwidth]{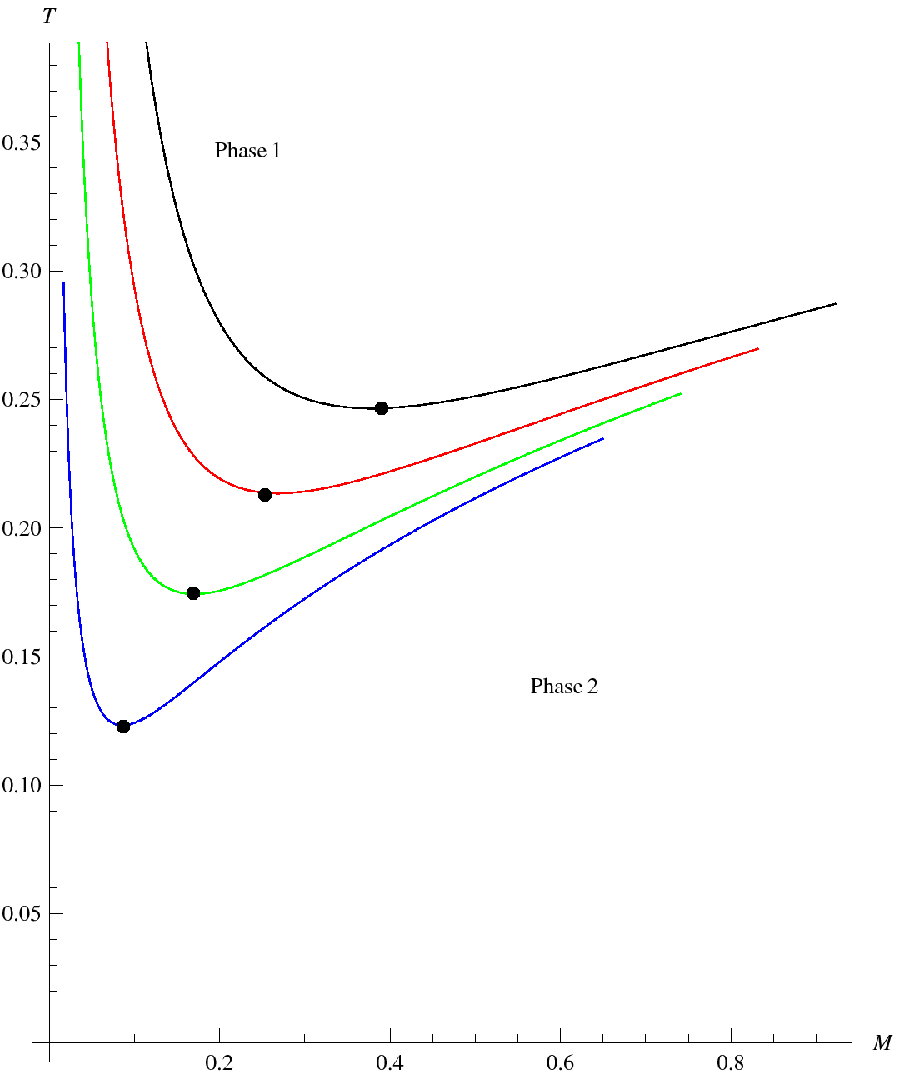} \caption{\label{f6}Temperature ($T$) vs. mass ($M$) for holographic screens
with constant $\Phi$, with $n=3$, $l=1$ and $\Phi^{2}=0.2$. The
values of $c$ for black, red, green, blue lines are 0, 0.2, 0.4,
0.6, respectively.}
\end{figure}

\begin{figure}[h]
\centering \includegraphics[width=0.4\textwidth]{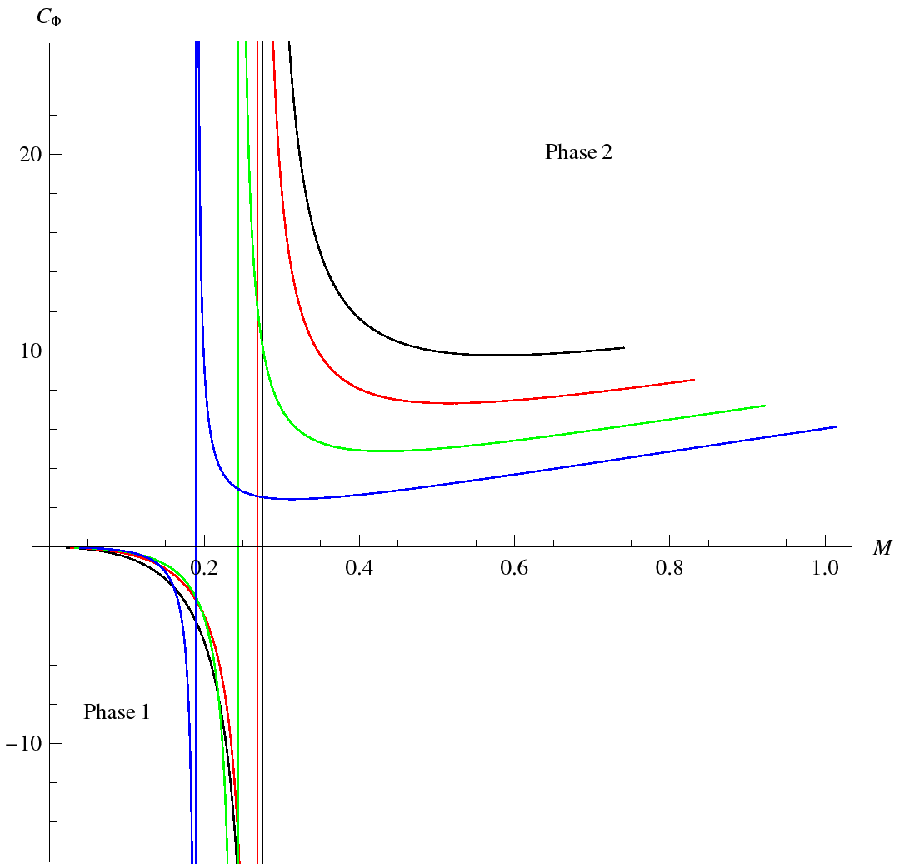} \caption{\label{f7}Heat capacity ($C_{\Phi}$) vs. mass ($M$) for holographic
screens with constant $\Phi$, with $n=3$, $l=1$ and $c=0.2$. The
values of $\Phi^{2}$ for black, red, green, blue lines are 0, 0.2,
0.4, 0.6, respectively.}
\end{figure}

\begin{figure}
\centering \includegraphics[width=0.4\textwidth]{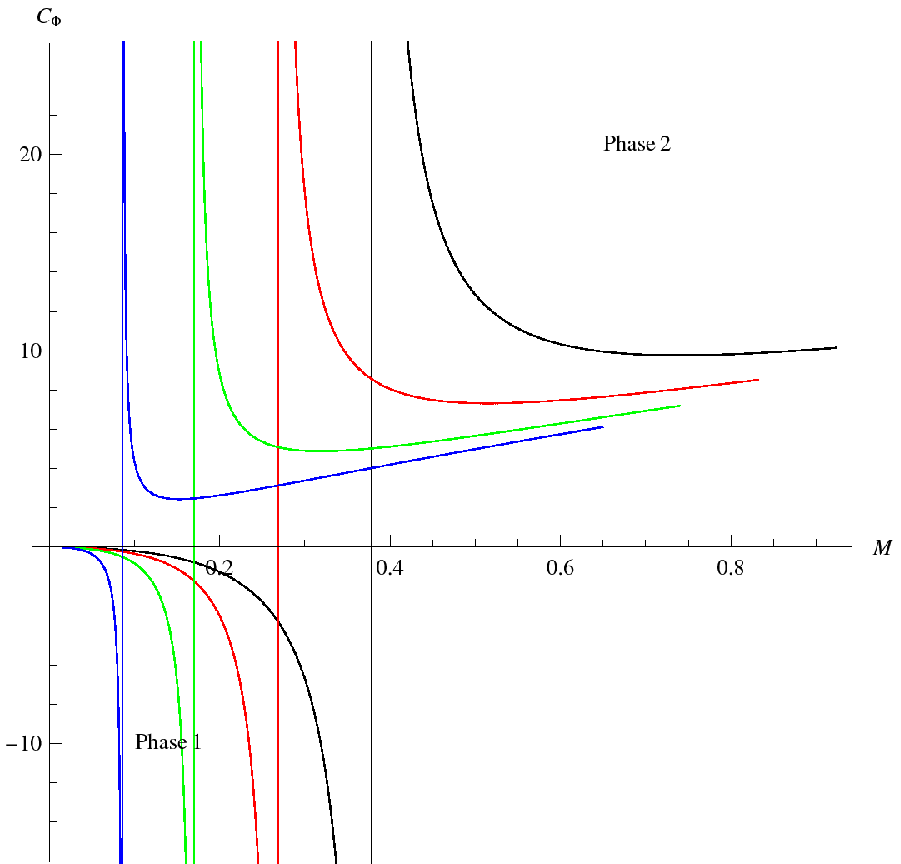} \caption{\label{f8}Heat capacity ($C_{\Phi}$) vs. mass ($M$) for holographic
screens with constant $\Phi$, with $n=3$, $l=1$ and $\Phi^{2}=0.2$.
The values of $c$ for black, red, green, blue lines are 0, 0.2, 0.4,
0.6, respectively.}
\end{figure}

Comparing Eq. (\ref{M_AdS}) with Eq. (\ref{M_RNAdS}), we see that
the constant $c$ and the constant $16\tilde{\Gamma}^{2}\Phi^{2}$
behave like a new combined constant $(c-16\tilde{\Gamma}^{2}\Phi^{2})$.
In addition, comparing Eqs. (\ref{T_AdS}), (\ref{C_AdS}) with Eqs.
(\ref{T_RNAdS}), (\ref{eq:C_RNAdS2}), we can see that the constant
$c$ and the constant $16\tilde{\Gamma}^{2}\Phi^{2}$ behave like
another new combined constant $(c+16\tilde{\Gamma}^{2}\Phi^{2})$.
Therefore, we illustrate the $T$-$M$, $C_{\Phi}$-$M$ diagram with
constant $c$ and varying $\Phi$ in Figs. \ref{f5}, \ref{f7}, and
with varying $c$ and constant $\Phi$ in Figs. \ref{f6}, \ref{f8},
respectively. Each holographic screen in Figs. \ref{f5}-\ref{f8}
undergoes a phase transition at the point $C_{\Phi}\rightarrow\infty$.
(We marked these turning points in Figs. \ref{f5}, \ref{f6}.)

Comparing Eqs. (\ref{T_AdS}), (\ref{C_AdS}) with Eqs. (\ref{T_RNAdS}),
(\ref{C_RNAdS1}), it is easy to see that the $C_{\Phi}$-$T$ diagram
of RN-AdS holographic screens will be similar to the $C$-$T$ diagram
of Schwarzschild AdS cases. Thus, we do not provide those diagrams
in this article. Although the $C_{\Phi}$-$T$ and $C$-$T$ diagrams
are similar, we should notice that there is a key deference between
the phase transitions in RN-AdS case with constant $\Phi$ and in
AdS case. When we change the mass of a holographic screen with specific
$c$ and $\Phi$, its radial parameter $r_{s}$ will change correspondingly.
Then we must alter its charge parameter $Q$ to ensure $\Phi=Q/4\tilde{\Gamma}r_{s}^{n-2}$
unchanged. Since holographic screens with different $c$ have different
$r_{s}$, we can not preserve all the electrostatic potentials of
these holographic screens in the same time. As a result, the curves
in Figs. \ref{f5} and \ref{f7} with different $c$ undergo separate
thermodynamical processes.

By setting the denominator in Eq. (\ref{eq:C_RNAdS2}) as 0, we get
$r_{s}$ at the phase transition point,

\begin{equation}
r_{pt}^{2}=\frac{(n-1)(n-2)(1-c-16\tilde{\Gamma}^{2}\Phi^{2})}{-2\Lambda}.\label{eq:rpt_ConstPhi}
\end{equation}

We should notice the following features in these Figs. There are two
phases of every holographic screen similar to Schwarzschild AdS case.
Also, when the holographic screen is in phase 1, it has smaller mass,
larger charge (not shown explicitly in the figures) and negative heat
capacity. In phase 2, it has larger mass, smaller charge (not shown
explicitly in the figures) and positive heat capacity. The phase transitions
are between phase 1 and phase 2, and at the phase transition point
the heat capacity divergent.

The difference from the Schwarzschild AdS case is, according to Eq.
(\ref{eq:C_RNAdS2}), that the existence of phase transition requires
$(c+16\tilde{\Gamma}^{2}\Phi^{2})<1$, in addition to require $c\geqslant0$.
It is to say that $0\le c<1$ and $0<\Phi<\sqrt{1-c}/4\tilde{\Gamma}$
or $0<\Phi<1/4\tilde{\Gamma}$ and $0\le c<(1-16\tilde{\Gamma}^{2}\Phi^{2})$.
It also shows that the holographic screens with $(c+16\tilde{\Gamma}^{2}\Phi^{2})>1$
have no phase transition, regardless of the respective values of $c$
or $\Phi$.

The temperature of the holographic screen at the transition point,
$T_{pt}$, can be deduced from Eq. (\ref{eq:rpt_ConstPhi}) and Eq.
(\ref{T_RNAdS}),

\begin{equation}
T_{pt}=\frac{1}{2\pi}\sqrt{\frac{-2\Lambda(n-2)(1-c-16\tilde{\Gamma}^{2}\Phi^{2})}{n-1}}.\label{eq:Tpt_RNAdS}
\end{equation}
When $0\le(c+16\tilde{\Gamma}^{2}\Phi^{2})<1$, the larger $(c+16\tilde{\Gamma}^{2}\Phi^{2})$
is, the lower $T_{pt}$ becomes. $T_{pt}$ is also the minimum temperature
of the corresponding holographic screen.

The mass of the holographic screen at the transition point, $M_{pt}$,
can be deduced from Eq. (\ref{eq:rpt_ConstPhi}) and Eq. (\ref{M_RNAdS}),

\begin{equation}
M_{pt}=\frac{1}{2}(1-c+16\tilde{\Gamma}^{2}\Phi^{2})r_{pt}^{n-2}-\frac{2\Lambda}{n(n-1)}r_{pt}^{n}.\label{eq:Mpt_RNAdS}
\end{equation}
When $0\le(c+16\tilde{\Gamma}^{2}\Phi^{2})<1$, and $c$, $\Phi$
become larger, the mass $M_{pt}$ will become smaller. 

Comparing Fig. \ref{f5} with Fig. \ref{f6}, we can see that the
varying $\Phi$ and the varying $c$ change the shape of the curve
in different ways. Given a small mass, the temperature discrepancies
of different $\Phi$-valued curves in Fig. \ref{f5} are smaller than
those of different $c$-valued curves in Fig. \ref{f6}. In contrary,
given a large mass, the temperature discrepancies of different $\Phi$-valued
curves in Fig. \ref{f5} are larger than those of different $c$-valued
curves in Fig. \ref{f6}.

Comparing Fig. \ref{f7} with Fig. \ref{f8}, it also shows that how
the varying $\Phi$ and the varying $c$ change the shape of the curve
in different ways. Given a small mass, the heat capacity discrepancies
of different $\Phi$-valued curves in Fig. \ref{f7} are smaller than
those of different $c$-valued curves in Fig. \ref{f8}. In contrary,
given a large mass, the heat capacity of different $\Phi$-valued
curves in Fig. \ref{f7} are larger than those of different $c$-valued
curves in Fig. \ref{f8}.

In \cite{Banerjee:2011raa,Banerjee:2011au}, the authors showed that
the phase transition of any (n+1) RN-AdS black hole with constant
$\Phi$ satisfies the Ehrenfest equations, so it is a second order
phase transition. This is the case for holographic screen with $c=0$.
We find that when $0\le(c+16\tilde{\Gamma}^{2}\Phi^{2})<1$, the thermodynamical
entities of holographic screen at the phase transition point also
satisfy the Ehrenfest equations, so the phase transitions with constant
$\Phi$ are also second order phase transitions. 

Following \cite{Banerjee:2011raa,Banerjee:2011au}, we first write
down the Ehrenfest equations for this system,

\begin{equation}
-(\frac{\partial\Phi}{\partial T})_{S}=\frac{C_{\Phi_{2}}-C_{\Phi_{1}}}{TQ(\alpha_{2}-\alpha_{1})}=\frac{\Delta C_{\Phi}}{TQ\Delta\alpha},\label{eq:Ehren1}
\end{equation}

\begin{equation}
-(\frac{\partial\Phi}{\partial T})_{Q}=\frac{\alpha_{2}-\alpha_{1}}{\kappa_{T_{2}}-\kappa_{T_{1}}}=\frac{\Delta\alpha}{\Delta\kappa_{T}}.\label{eq:Ehren2}
\end{equation}
here,$\alpha=\frac{1}{Q}(\frac{\partial Q}{\partial T})_{\Phi}$,
$\kappa_{T}=\frac{1}{Q}(\frac{\partial Q}{\partial\Phi})_{T}$, which
are similar to the volume expansion coefficient and the isothermal
compressibility in the standard thermodynamics respectively. 

By noting $Q=Q(\Phi,S)$ and using Eq. (\ref{C_RNAdS1}), we get

\begin{equation}
\alpha=\frac{1}{Q}(\frac{\partial Q}{\partial T})_{\Phi}=\frac{1}{Q}(\frac{\partial Q}{\partial S})_{\Phi}(\frac{\partial S}{\partial T})_{\Phi}=\frac{C_{\Phi}}{QT}(\frac{\partial Q}{\partial S})_{\Phi}.\label{eq:Qalpha0}
\end{equation}
In the same way, by noting $Q=Q(\Phi,T)$, we get
\begin{equation}
\kappa_{T}=\frac{1}{Q}(\frac{\partial Q}{\partial\Phi})_{T}=-\frac{1}{Q}(\frac{\partial T}{\partial\Phi})_{Q}/(\frac{\partial T}{\partial Q})_{\Phi}=-\alpha(\frac{\partial T}{\partial\Phi})_{Q}.\label{eq:Qkappa0}
\end{equation}
At the phase transition point, $C_{\phi}$ diverges. From Eq. (\ref{eq:Qalpha0})
and Eq. (\ref{eq:Qkappa0}), we can see $\alpha$ and $\kappa_{T}$
diverge too. Or calculate $\alpha$ and $\kappa_{T}$ explicitly,

\begin{equation}
\alpha=\frac{4\pi(n-2)r_{s}}{-(n-2)(1-c-16\tilde{\Gamma}^{2}\Phi^{2})-\frac{2\Lambda}{n-1}r_{s}^{2}},\label{eq:Qalpha}
\end{equation}

\begin{equation}
\kappa_{T}=\frac{-(n-2)[1-c-16(2n-3)\tilde{\Gamma}^{2}\Phi^{2}]-\frac{2\Lambda}{n-1}r_{s}^{2}}{-(n-2)(1-c-16\tilde{\Gamma}^{2}\Phi^{2})-\frac{2\Lambda}{n-1}r_{s}^{2}}.\label{eq:Qkappa}
\end{equation}
We observe $\alpha$ and $\kappa_{T}$ have the same denominator as
$C_{\Phi}$. Of course, they all diverge at the same point. But when
we calculate the R.H.S. of the Ehrenfest equations Eq. (\ref{eq:Ehren1})
and Eq. (\ref{eq:Ehren2}), the results will be finite. If we represent
$C_{\Phi}$ and $\alpha$ as $\frac{f(r)}{g(r)}$ and $\frac{h(r)}{g(r)}$
respectively, note $g(r)\bigg|{}_{r\rightarrow r_{pt}}=0$, we have
\begin{align*}
\frac{\Delta C_{\Phi}}{\Delta\alpha} & =\left(\frac{\frac{f(r_{2})}{g(r_{2})}-\frac{f(r_{1})}{g(r_{1})}}{\frac{h(r_{2})}{g(r_{2})}-\frac{h(r_{1})}{g(r_{1})}}\right)\bigg|{}_{r_{1},r_{2}\rightarrow r_{pt}}=\frac{f(r_{pt})}{h(r_{pt})}
\end{align*}
So from Eq. (\ref{eq:Qalpha0}), Eq. (\ref{S}) and Eq. (\ref{Phi}),
the R.H.S. of the first Ehrenfest equation Eq. (\ref{eq:Ehren1})
becomes

\begin{equation}
\frac{\Delta C_{\Phi}}{TQ\Delta\alpha}=[(\frac{\partial S}{\partial Q})_{\Phi}]_{pt}=\left(\frac{\pi r_{s}}{8(n-2)\tilde{\Gamma}^{2}\Phi}\right)_{pt}.\label{eq:RHSEhren1}
\end{equation}
In the same way, using Eq. (\ref{eq:Qkappa0}), the R.H.S. of the
second Ehrenfest equation Eq. (\ref{eq:Ehren2}) becomes
\begin{equation}
\frac{\Delta\alpha}{\Delta\kappa_{T}}=-[(\frac{\partial\Phi}{\partial T})_{Q}]_{pt}.\label{eq:RHSEhren2}
\end{equation}

Calculating the L.H.S. of Eq. (\ref{eq:Ehren1}) straightforwardly,
we get 

\begin{equation}
-(\frac{\partial\Phi}{\partial T})_{S}=\frac{\pi r_{s}}{8\tilde{\Gamma}^{2}(n-2)\Phi}.\label{eq:LHSEhren1}
\end{equation}
From Eq. (\ref{eq:LHSEhren1}) and Eq. (\ref{eq:RHSEhren1}), we see
the first Ehrenfest equation is satisfied. And Eq.(\ref{eq:RHSEhren2})
is already the same as the second Ehrenfest equation.

For a consistency check, we should verify that the Prigogine-Defay
ratio $\Pi$, which is the ratio of the R.H.S. of the first and the
second Ehrenfest equations, is equal to 1. By means of the straightforward
calculation the R.H.S. of Eq.(\ref{eq:RHSEhren2}) , we obtain

\[
-(\frac{\partial\Phi}{\partial T})_{Q}=\frac{4\pi r_{s}\Phi}{-(1-c)+16(2n-3)\tilde{\Gamma}^{2}\Phi^{2}-\frac{2\Lambda}{(n-1)(n-2)}r_{s}^{2}},
\]
use $r_{pt}$ in Eq. (\ref{eq:rpt_ConstPhi}), we have

\begin{equation}
-[(\frac{\partial\Phi}{\partial T})_{Q}]_{pt}=\frac{\pi r_{pt}}{8(n-2)\tilde{\Gamma}^{2}\Phi}.\label{eq:LHSEhren2}
\end{equation}
Notice this is equal to Eq. (\ref{eq:RHSEhren1}), so $\Pi=1$ is
verified.

\subsection{Phase transition with constant $Q$}

In RN-AdS spacetime, the phase transition of holographic screen with
constant $Q$ is more complicated. The mass and temperature are showed
in Eq. (\ref{M}) and Eq. (\ref{T}). In addition, the heat capacity
with constant $Q$ is, 
\begin{eqnarray}
C_{Q} & = & (\frac{\partial M}{\partial T})_{Q}=\frac{\pi r_{s}^{n-1}}{2\tilde{\Gamma}}\frac{(n-2)(1-c)r_{s}^{2n-4}-\frac{2\Lambda}{n-1}r_{s}^{2n-2}-(n-2)Q^{2}}{-(n-2)(1-c)r_{s}^{2n-4}-\frac{2\Lambda}{n-1}r_{s}^{2n-2}+(n-2)(2n-3)Q^{2}}.\label{eq:CQ_RNAdS}
\end{eqnarray}

To analyze the phase structure, we should set the denominator of $C_{Q}$
equal to 0, and solve for $r_{pt}$. It is easy to see that if $c>1$
the denominator of $C_{Q}$ is definitely positive. So for the holographic
screens with $c>1$, there is no phase transition at all.

For simplicity, we set $n=3$,$l=1$ in following discussion. $M$,
$T$ and $C_{Q}$ become,

\begin{equation}
M=\frac{1}{2}\left[(1-c)r_{s}+\frac{Q^{2}}{r_{s}}+r_{s}^{3}\right],\label{eq:M3}
\end{equation}

\begin{equation}
T=\frac{1}{4\pi r_{s}}(1-c-\frac{Q^{2}}{r_{s}^{2}}+3r_{s}^{2}),\label{eq:T3}
\end{equation}

\begin{equation}
C_{Q}=2\pi r_{s}^{2}\frac{(1-c)r_{s}^{2}+3r_{s}^{4}-Q^{2}}{-(1-c)r_{s}^{2}+3r_{s}^{4}+3Q^{2}}.\label{eq:CQ3}
\end{equation}

\begin{figure}[h]
\centering \includegraphics[width=0.4\textwidth]{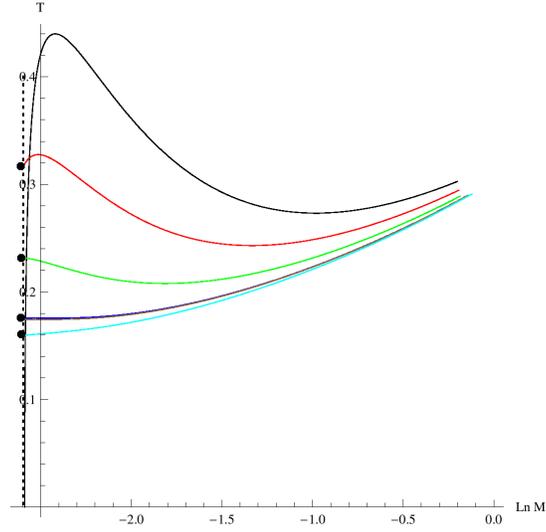}
\caption{\label{f9}Temperature ($T$) vs. logarithm of mass ($\ln M$) for
holographic screens with constant $Q$, with $n=3$, $l=1$ and $Q=0.075$.
The values of $c$ for the black, red, green, blue, brown, cyan lines
are 0, 0.2, 0.4, 0.545, 0.55, 0.6, respectively.}
\end{figure}

\begin{figure}[h]
\centering \includegraphics[width=0.4\textwidth]{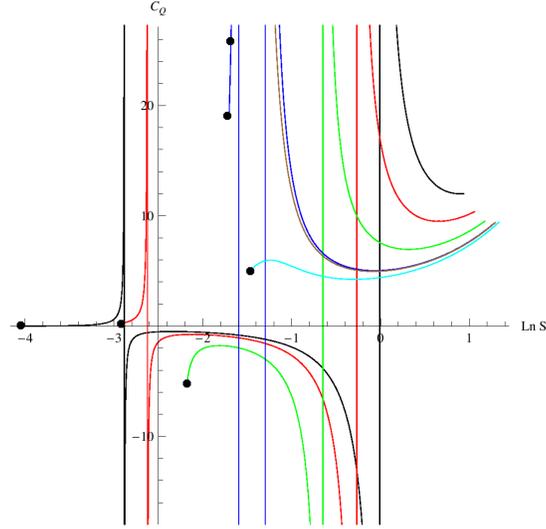}
\caption{\label{f10}Heat capacity $C_{Q}$ vs. logarithm of entropy ($\ln S$)
for holographic screens with constant $Q$, with $n=3$, $l=1$ and
$Q=0.075$. The values of $c$ for the black, red, green, blue, brown,
cyan lines are 0, 0.2, 0.4, 0.545, 0.55, 0.6, respectively.}
\end{figure}

In Figs. \ref{f9}, \ref{f10}, we set $Q=0.075$, and the values
of $c$ for the black, red, green, blue, brown, cyan lines are 0,
0.2, 0.4, 0.545, 0.55, 0.6, respectively. These numbers are elabrately
selected, the reason will be explianed below. In order to shorten
the horizontal coordinate, we use the logarithmic coordinate. We find
several new features of the phase transitions with constant $Q$. 

The curves in Figs. \ref{f9}, \ref{f10} have cut-offs on the left,
which are labeled by black dots. The cut-offs are determined by the
extremal limit of RN-AdS black hole. When $c=0$, the black hole temperature
must be positive, so there is a lower-limit of the black hole mass
$M_{min}$, 

\begin{eqnarray}
M_{min} & = & \frac{1}{2}\sqrt{\frac{\sqrt{12Q^{2}+1}-1}{6}}(1+\frac{6Q^{2}}{\sqrt{12Q^{2}+1}-1}+\frac{\sqrt{12Q^{2}+1}-1}{6}).\label{M_min}
\end{eqnarray}
This is also the lower-limit of all the other holographic screens.

In addition, it is possibly for the denominator of $C_{Q}$ to have
two, one or zero positive roots, which consequently makes it possible
for holographic screen to have two, one or zero phase transition points.

\begin{figure}
\centering{}\includegraphics[width=0.4\paperwidth]{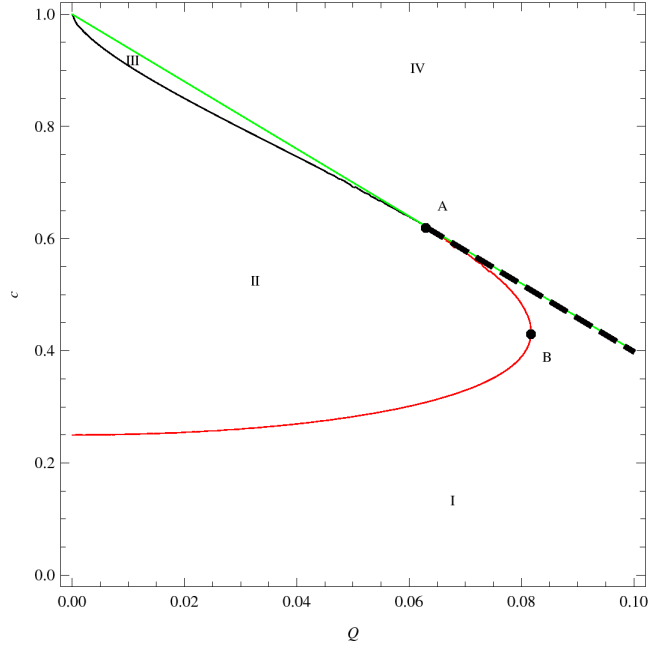}\caption{\label{fig:11}Values of $Q$ and $c$ for different phase structures.
The region I and II represent the holographic screens can undergo
two and one phase transitions, respectivlly. The region III and IV
represent the holographic screens can undergo no transition, but for
different reasons. The dashed line stands for the holographic screens
have critical points.}
\end{figure}

These are the two reasons which make the phase structure of holographic
screen intricate. In Fig. \ref{fig:11}, we present several regions
in the $Q-c$ plane. The holographic screens, with values of $Q$
and $c$ in different regions, have different phase structures. The
region I, II, III are under the straight green line in Fig. \ref{fig:11},
which is the region of $c<1-6Q$. In these regions, from the divergence
points of $C_{Q}$, we obtain two $r_{pt}$. Then substituting these
$r_{pt}$ in Eq. (\ref{eq:M3}) we get two $M_{pt}$. The region I
represents these two $M_{pt}$ are both bigger than $M_{min}$. So
the corresponding holographic screens can undergo both the two phase
transitions. The region II stands for that $M_{min}$ lies between
these two $M_{pt}$, consequently the holographic screens in this
region can only undergo one phase transition. The region III represents
these two $M_{pt}$ are both smaller than $M_{min}$, hence the holographic
screen in this region can not undergo any phase transition. In the
region IV, which indicates $c>1-6Q$, $C_{Q}$ has no divergent point
at all, so the corresponding holographic screens have only one phase.
We should note that when $0.063<Q<0.081$, which means the dot $(Q,c)$
in Fig. \ref{fig:11} lies between two vertical lines through the
dots A and B, the holographic screens with various $c$ have different
phase structures. This is why we choose $Q=0.075$ and the specific
values of $c$ in Fig. \ref{f10}. 

In Fig. \ref{f10}, the black curve ($c=0$), the red curve ($c=0.2$)
and the blue curve ($c=0.545$) have 3 phases. The heat capacity in
the left is positive, in the middle it is negative (For for the blue
curve, the negative heat capacity is too small, so it is not contained
in Fig.\ref{f10}. ), and in the right it is positive again. So there
are two phase transition points for these holographic screens. The
green curve ($c=0.4$) has only 2 phases, because of the $M_{min}$
cut-off, the heat capacity in the left is negative and in the right
is positive. The cyan curve ($c=0.6$) has only one phase, the heat
capacity is always positive.

In Fig. \ref{f10}, the brown curve $c=0.55$ have 2 phases, (the
left part of this curve is shown as coinside with the blue curve ($c=0.545$),
starting from the upper dot.) and the heat capacities of both phases
are positive. The point, at which $C_{Q}$ is divergent, is the critical
point of the holographic screen. The condition for the existence of
critical point is that the denominator of Eq. (\ref{eq:CQ_RNAdS})
has two equal real roots. Thus we get the general value of critical
charge,

\begin{equation}
Q_{c}^{2}=(\frac{(n-2)^{2}}{-2\Lambda})^{n-2}\frac{(1-c)^{n-1}}{(n-1)(2n-3)}.\label{Q_critical}
\end{equation}
And setting $n=3$ and $l=1$ in Eq. (\ref{Q_critical}), we get 
\begin{equation}
Q_{c}^{2}=\frac{(1-c)^{2}}{36}.\label{eq:Qc3}
\end{equation}
We note this is the straight green line in Fig. \ref{fig:11}. By
setting the denominator of $C_{Q}$ in Eq. (\ref{eq:CQ3}) equal to
0, we can solve the critical radial parameter $r_{c}$, and then substitute
$r_{c}$ in Eq. (\ref{eq:M3}), we obtain the critical mass $M_{c}$,

\begin{equation}
M_{c}=\frac{1}{3}\sqrt{\frac{2}{3}}(1-c)^{3/2}.\label{eq:Mc}
\end{equation}
From Eq. (\ref{M_min}) we can find the $M_{min}$ for $Q_{c}$,
\begin{equation}
M_{min_{c}}=\frac{-3+(1-c)^{2}+\sqrt{3}\sqrt{3+(1-c)^{2}}}{9\sqrt{-6+2\sqrt{3}\sqrt{3+(1-c)^{2}}}}.\label{eq:Mminc}
\end{equation}
So only the holographic screens, which satisfy $M_{c}<M_{min_{c}}$,
will undergo the critical point. The condition $M_{c}<M_{min_{c}}$
is equal to $c<0.623$. And in Fig. \ref{fig:11}, this condition
means the part of the straight line at the right of dot A, which is
marked by dashed line.

\section{Conclusion}

The thermodynamical properties of holographic screen play an important
role in Verlinde's entropic force scenario. According to previous
works, it is believed that the holographic screen has its energy,
temperature and entropy. There is also a corresponding first law among
these thermodynamical variables. Although thermodynamics of holographic
screens origins from its analogy with black hole thermodynamics, it
has some pretty unique properties. Naturally, the next step is to
examine whether there are some interesting phase transition properties
of holographic screens. In this article, we analyzed the general case
of holographic screens enclosing a RN-AdS black hole. Starting from
the $(n+1)-$dimensional metric, we emphasized the equivalence between
the integrated form of first law and the $(r-r)$ component of the
Einstein equations in the neighborhood of holographic screen. A notable
feature of the holographic screen's phase structure is its dependence
on the Newton potential $c$. We found that in different cases, phase
transitions only exist when $c$ picks some specific range of values. 

The simplest case is $Q=0$, while the metric degenerates to Schwarzschild-AdS
spacetime. By examining the phase diagram for fixed values of $c$,
we found that the states of holographic screen are separated into
two phases by the divergent point of the heat capacity $C$. The phase
with smaller mass and entropy has a negative heat capacity, so it
is unstable. The other phase with larger mass and entropy is stable.
For large values of $c$, the phase transition properties are more
significant. However, the phase transitions only exist for $0\leqslant c<1$. 

The charged holographic screen has another parameter, $Q$ or $\Phi$.
Firstly we considered the case of constant $\Phi$. For fixed values
of $c$ and $\Phi$, there are two phases. The phase structure is
rather similar with the case of $Q=0$. The difference here is that
the curves in phase diagrams are affected not only by the Newton potential
$c$ but also by the electrostatic potential $\Phi$. Varying the
values of $c$ and $\Phi$ changes the shapes of curves in different
ways. The phase transition only exist for $0\leqslant c<1$ and $0<\Phi<\frac{1}{4\tilde{\Gamma}}$.
One should notice there are no phase transition for holographic screens
with $\left(c+16\tilde{\Gamma}^{2}\Phi^{2}\right)>1$, regardless
of the values of $\Phi$. By employing the Ehrenfest's scheme, we
find that the phase transition of holographic screen here is second
order. 

The case of holographic phase transitions at constant $Q$ are a bit
more complicated. With the parameter $Q$ in the RN-AdS metric, the
black hole has an extreme limit. Since the temperature of RN-AdS black
hole cannot be negative, there is a lower limit of the mass parameter.
A consequent implication is that the mass parameter of holographic
screens enclosing an RN-AdS black hole also has this lower limit.
This property of RN-AdS black hole restricts the definition region
of the radius of holographic screens. As a consequence, with specific
values of $c$ and $Q$, there is a starting point for each curve
in the phase diagram. This is an important characteristic for the
case of constant $Q$. According to these constraints, for different
values of $c$ and $Q$, the holographic screens can have two, one
or zero phase transition points. We gave the definite regions of $Q$
and $c$ for each phase structure. For some specific values of $c$
and $Q$, the holographic screens have critical points. We also gave
the corresponding values of $Q$ and $c$ for the existence of critical
points. One other thing to note is that there are no such phenomenons
for the case of constant $\Phi$. Because the electrostatic potential
$\Phi$ depends on $Q$ and $r_{s}$, for fixed $\Phi$, smaller $r_{s}$
corresponds to smaller $Q$, and there is no lower limit for $r_{s}$
and $T$ other than zero. Due to the complexity, we only gave some
numerical results for $n=3$, $l=1$. 

The above discussions show that there are many interesting properties
with the phase transition of holographic screens. We believe that
our analysis will be useful for understanding the thermodynamic properties
of holographic screens, as well as the entropic force scenario.

\acknowledgments

We thank K.N. Shao for useful discussions. This work is supported
in part by the NSF of China Grant No. 11105118, No. 10775116, No.
11075138, and 973-Program Grant No. 2005CB724508.

\appendix
\bibliographystyle{apsrev4-1}
\bibliography{hsphasetransitions}
 
\end{document}